\documentclass[twocolumn]{article}
\pdfoutput=1
\usepackage[pdftex]{graphicx} \pdfcompresslevel=9
\usepackage{hyperref}

\usepackage{cite}

\usepackage{color}
\usepackage{amssymb}
\usepackage{amsmath}
\usepackage[utf8x]{inputenc}
\usepackage{wrapfig}

\usepackage[ruled,vlined]{algorithm2e}

\usepackage{float}

\makeatletter
\g@addto@macro\@maketitle{%
\par
\noindent
\begingroup
\def\@captype{figure}
\begin{minipage}[!t]{0.48\linewidth}
  \includegraphics[width=\linewidth]{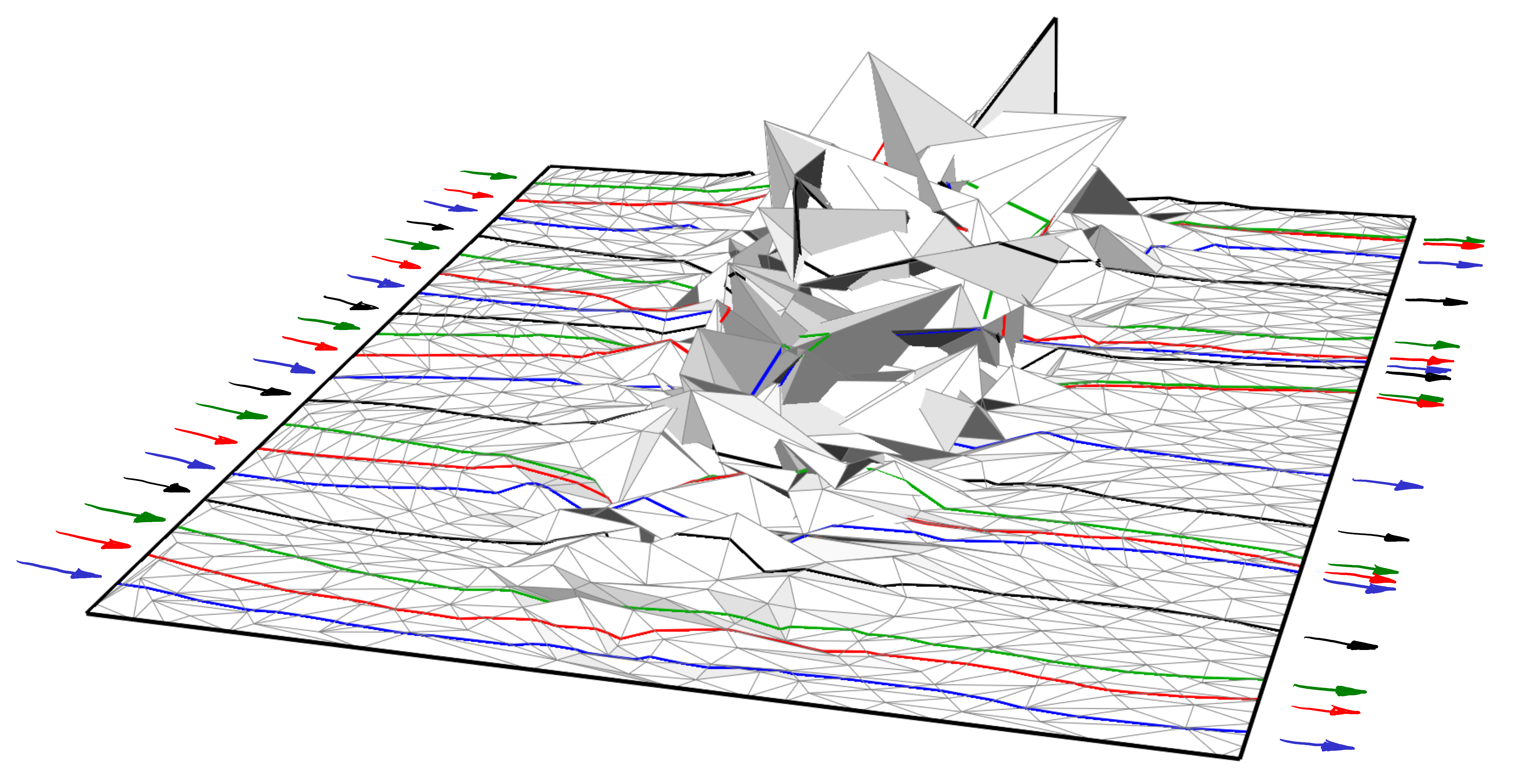}
\end{minipage}
\hspace{.5cm}
\begin{minipage}[!t]{0.48\linewidth}
 \includegraphics[width=.48\linewidth]{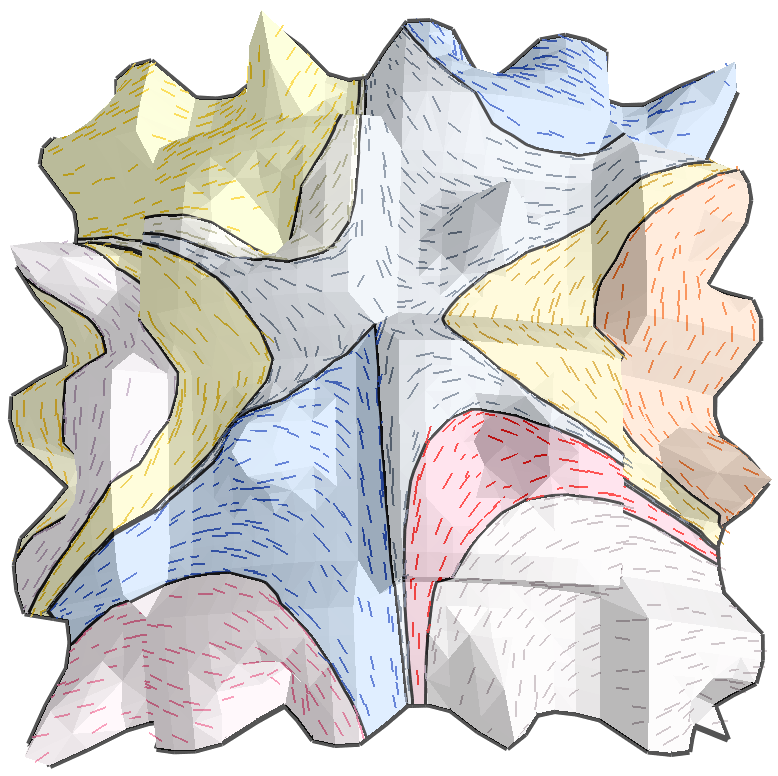}
 \includegraphics[width=.48\linewidth]{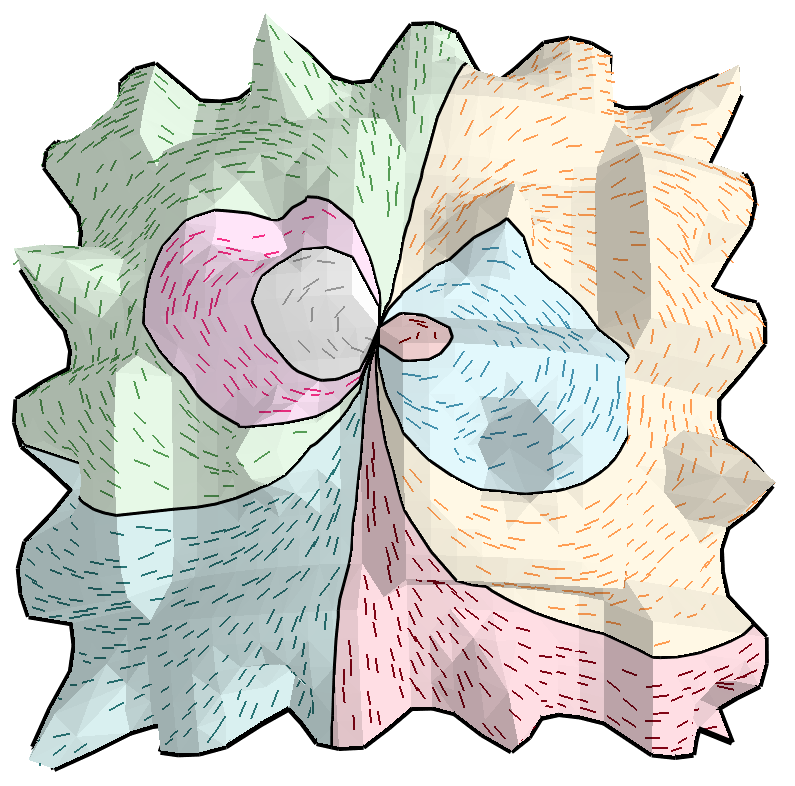}
\end{minipage}
\caption{Our algorithm traces polylines on triangulated surfaces. Unlike streamline tracing algorithms, polylines produced by our technique cannot cross each others. It works even with highly perturbed surfaces (left) and supports any type of vector field singularities (right). This property is required to segment surfaces with chart boundaries aligned with a vector field (right).}
  \label{fig:teaser}
\endgroup
\par
\vspace{2\baselineskip}%
}
\makeatother

\title{Tracing cross-free polylines oriented by a N-symmetry direction field on triangulated surfaces}

\author{N. Ray\thanks{e-mail: ray@loria.fr}\\INRIA  \and D. Sokolov\thanks{e-mail: sokolovd@loria.fr}\\Universit\'e de Lorraine}

\begin{document}

\maketitle

\begin{abstract}

We propose an algorithm for tracing polylines on a triangle mesh such that: they are aligned with a N-symmetry direction field, and two such polylines cannot cross or merge. This property is fundamental for mesh segmentation and is very difficult to enforce with numerical integration of vector fields. We propose an alternative solution based on ``stream-mesh'', a new combinatorial data structure that defines, for each point of a triangle edge, where the corresponding polyline leaves the triangle. It makes it possible to trace polylines by iteratively crossing triangles. Vector field singularities and polyline/vertex crossing are characterized and consistently handled. The polylines inherits the cross-free property of the stream-mesh, except inside triangles where avoiding local overlaps would require higher order polycurves.

%\begin{classification} % according to http://www.acm.org/class/1998/
%\CCScat{Computer Graphics}{I.3.3}{Picture/Image Generation}{Line and curve generation}
%\end{classification}

\end{abstract}

\section{Introduction}

Segmentation of triangulated surfaces that aligns chart boundaries with a vector field (or, more generally, a direction field) often exhibits useful properties for computer graphics applications. For instance, alignment with the main curvature directions allows for quad dominant remeshing \cite{Alliez2003}, following the gradient of a scalar field allows to compute pure quad decomposition using Morse-Smale complexes \cite{garland2006,eugene2012}, and streamlines of a cross field can decompose a mesh into quad shaped domains \cite{Kowalski2013}.

\begin{figure}[t]
\centerline{
\includegraphics[height=1.8cm,angle=90]{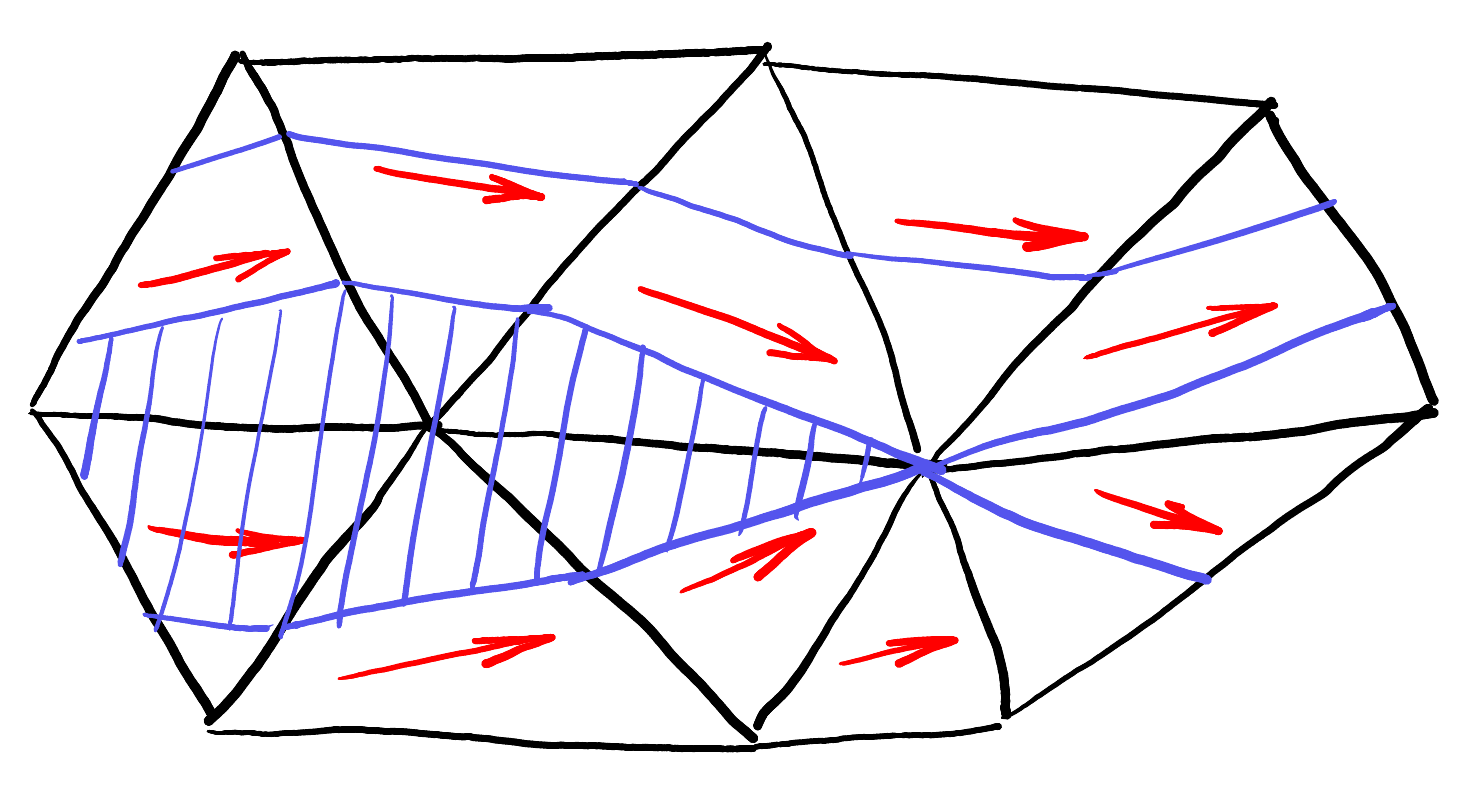}
\hspace{.2cm}
\raisebox{.6cm}{\includegraphics[width=5cm]{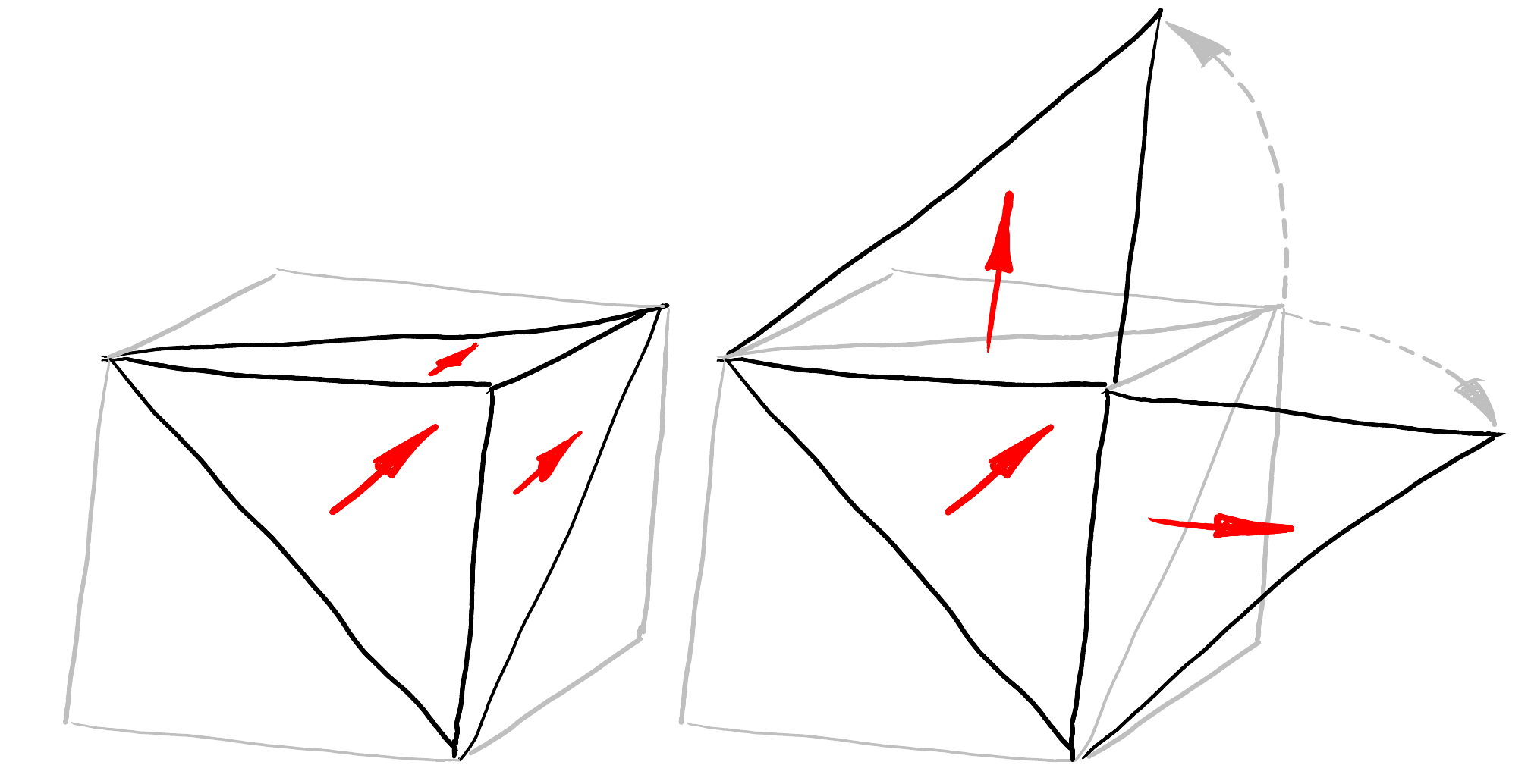}}
}
\hspace{.9cm}{\textbf{ (a)\hspace{3.4cm} (b)}}
\caption{\textbf{(a)} all streamlines from the dashed area merge on an edge, and split on a vertex. \textbf{(b)} constant per triangle tangent vector fields have direction discontinuities along edges due to vertex angle defect.}
  \label{fig:constflow}
\end{figure}

The simplest solution to trace such polylines is to define a constant per triangle vector field. However, in this representation, streamlines will have merging, splittings and crossings, as illustrated in Figure~\ref{fig:constflow}(a) and further detailed in \cite{eugene2012}. As explained in Section \ref{sec:field}, only Zhang \emph{et al}'s vector field representation \cite{gre2006} allows to completely avoid these issues, but computing accurate enough streamlines is still very difficult.

\begin{figure*}[t]
\centerline{\includegraphics[width=\linewidth]{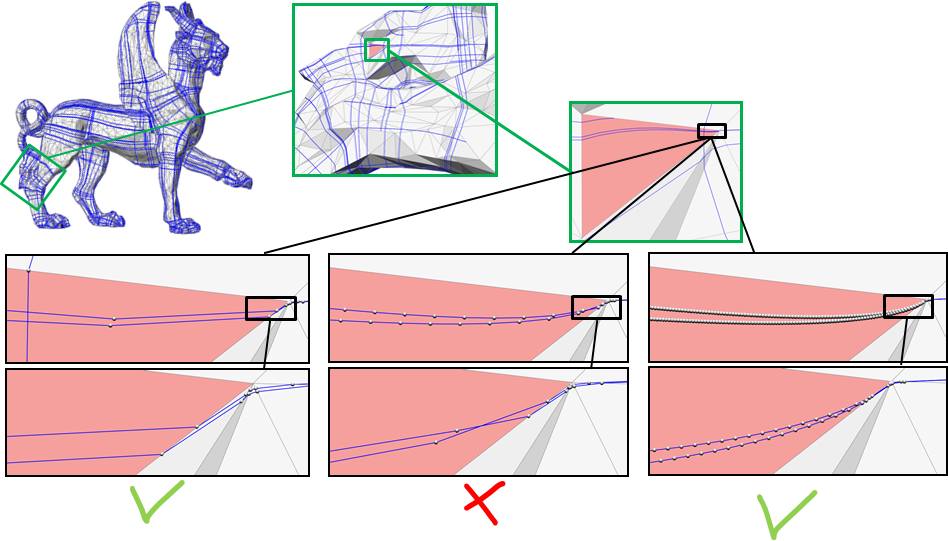}}
\caption{\textbf{Numerical integration issues.} Streamlines are traced on a direction field using an order 4 Runge-Kutta with integration step length of (from left to right) $1/10$, $1/100$ and $1/1000$ times the average edge length. A crossing appeared on the pink triangle with $1/10$ and $1/100$ step length.}  
  \label{fig:grecompare}
\end{figure*}

The continuous counter part of our problem would be to trace streamlines of a vector field tangent to a smooth surface. As continuous streamlines do not cross, an accurate enough numerical field integration will share this property. The representation of Zhang \emph{et al} provides an equivalent smooth problem without having to refine the field and the surface. We can therefore characterize the integration accuracy only by the step length of the order four Runge-Kutta field integration algorithm. In practice, the field is derived from the surface geometry, making it common to have streamlines very close to each other (local symmetries, almost aligned singularities, etc.). We observed (Figure~\ref{fig:grecompare}) that with a integration step length higher than equals to $1/100$ times the average edge length, most models have at least one streamline crossing. From a practical point of view, an accuracy that would prevents most conflicts ($\approx 1/1000$ times the average edge length) is more than $100$ times slower than our algorithm. Moreover, it would be very difficult to refine the integration because: tracing a streamline is a basic operation of the segmentation process and removing a crossing requires to refine both streamlines, and higher accuracy may produce new crossings.

Our solution is much faster than numerical field integration (crossing a triangle takes approximately the time to perform $10$ RK4 steps), and allows to prevent crossings without tuning any ``accuracy'' parameters. Those benefits come at the expense of the fitting quality with the input field. For mesh segmentation, where the input field only give a coarse approximation of the desired edge direction \cite{PGP,tmesh,Kowalski2013}, it is more important to have a guaranty that streamlines will not cross than a better fitting with the input field.

An overview of our method is presented in Algorithm~\ref{alg:overview} and illustrated in Figure~\ref{fig:crossmesh}. For each crossed triangle we construct on the fly a stream-mesh structure which is essentially the original triangle split in a way that each stream-halfedge can be flagged as input/output/tangent.
Each vertex of the polyline is defined by its underlying mesh halfedge $e$, and its barycentric coordinate $c$ (and $1-c$) on this halfedge. A similar representation $e_{sm},c_{sm}$ is used for crossing the stream-mesh. A streamline finish either when the halfedge has no associated facet (mesh boundary), or when $c \in [1\dots 2]$ instead of being in $[0\dots 1]$ to denote an output on a singular vertex e.g. a sink of the vector field.

\textbf{Limitation} Having a single polyline segment inside each triangle may produce degenerated geometries: local (limited to a single edge) overlaps and regions of triangle that are not covered by any polyline (inset figure of section~\ref{sec:tracingalgo}).

\begin{algorithm}
\SetAlgoLined
\KwOut{Polyline $PL$}
\KwIn{$PL$ extremity halfedge $e$}
\KwIn{$PL$ extremity position $c$ on $e$}
\While {e.has\_facet() AND $c \in [0\dots 1]$}{
	$sm \leftarrow stream\_mesh(e.facet())$ (Section~\ref{sec:algorithm})\;
	$e_{sm},c_{sm} \leftarrow sm.import\_position(e,c)$\;
	\While {$e_{sm}$.has\_facet()}{
		$e_{sm},c_{sm} \leftarrow cross\_facet(e_{sm},c_{sm})$(Section~\ref{sec:tracingalgo})\;
	}
	$e,c \leftarrow sm.export\_position(e_{sm},c_{sm})$\;
	$PL.add\_vertex(position(e,c)))$\;
}
\caption{Algorithm overview}
\label{alg:overview}
\end{algorithm}

\subsection*{Previous Works}

To the best of our knowledge, no prior work directly addresses our problem. However, it is interesting to consider solutions developed for $2D$ streamline tracing, to notice similar issues occurring for tracing other types of curves on surfaces, and to give an overview of the tangent vector field and, more generally, N-symmetry direction field design algorithms.

\subsubsection*{Streamline tracing}

Tracing streamlines of $2D$ or $3D$ vector fields is a common task \cite{SpencerLCZ09,Rossl2012} in visualization. In most cases, an order four Runge Kutta (RK4) integration scheme performs well. For piecewise linear vector field on a triangulation, a more robust solution \cite{BhatiaJBCLNP11} was proposed: an edge map directly matches in/out flow intervals of the triangle border. However, they assume the field is defined on vertices and linearly interpolated inside each triangle, and therefore cannot address the vertex angle defect issue (discussed in Section ~\ref{sec:field}). 

\begin{figure}[t]
\centerline{\includegraphics[width=\linewidth]{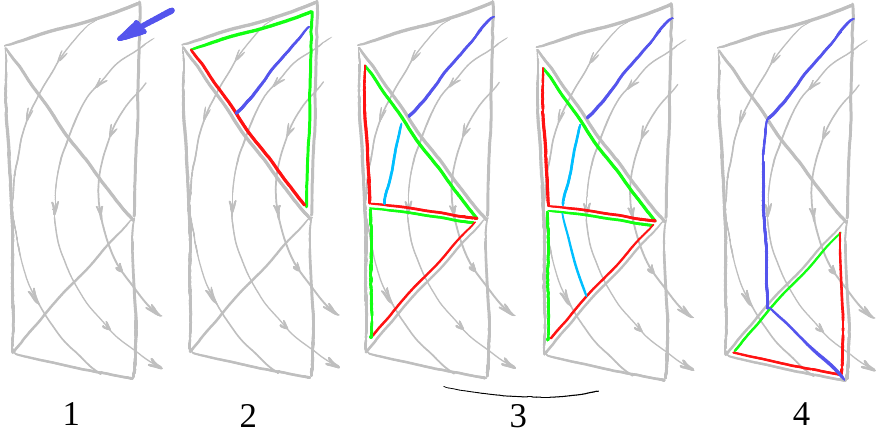}}
\caption{At each step, the algorithm crosses a triangle: a stream-mesh is constructed (green and red halfedges), and the polyline (blue) crosses the stream-mesh. In step 3, the polyline crosses two stream-faces, but the result is a single segment crossing the triangle (visible in step 4).}    
  \label{fig:crossmesh}
\end{figure}

\subsubsection*{Tracing curves on triangulated surfaces}

Tracing curves on triangulated surfaces is a challenging task because the curve may cross triangles, follow edges, and pass through vertices \cite{li:MEE:2005}. All such configurations are naturally managed by our representation: a polyline passing through a vertex is considered as crossing a subset of its adjacent triangles, with all vertices of the polyline located on the vertex of the surface.

For computing optimal systems of loops \cite{cl-oslos-05}, one needs to distinguish the order between curves following the same edge, leading to a complex data structure where all curves following the same edge need to be ordered. Special efforts \cite{Martinez2005,Surazhsky2005,Polthier:2006:SGP:1185657.1185664} have also been devoted to tracing geodesics where, as in our case, the angle defect plays an important role.

Recent works \cite{eugene2012} compute Morse decomposition of piecewise constant vector fields by converting them into a combinatorial structure. It results in a robust algorithm, but the streamlines (edges of the Morse complex) merges due to the input field.

\subsubsection*{Direction field design}

Many algorithms \cite{gre2006,Wang:2006:ESS:1141911.1141991,Fisher07designof} allow to design tangent vector fields. The produced field can be continuous enough to have (continuous) streamlines that does not cross each others \cite{gre2006}, eventually at the expense of simultaneously refining the surface \cite{Wang:2006:ESS:1141911.1141991}.

For mesh segmentation, it is more common to use N-symmetry direction fields than tangent vector fields but, as pointed by \cite{quadcover}: an N-symmetry direction field is equivalent to a vector field on an N-covering of the surface. Such fields were used for quad remeshing based on global parameterization \cite{PGP}. The lack of control over these direction fields topology was addressed later \cite{nrosy,NSDF}. A common representation \cite{NSDF,DFD,Bommes:2009,quadcover} samples the direction on triangles, and explicits the field rotation between adjacent triangles.

\section{Field representation}
\label{sec:field}

In many cases, it is impossible to trace cross-free streamlines due to the N-symmetry direction field representation. We detail this issue and introduce an alternative representation. 

Most representations of tangent vector fields are polynomial on each triangle. These vector fields are differentiable everywhere on each triangle, so their direction expressed as an angle in a local basis of the triangle is also differentiable everywhere except where the vector field is null.

This continuity of the field on triangles also involves discontinuities of the field direction on edges in the vicinity of vertices with non zero angle defect. Indeed, along an infinitesimal circle around the vertex, a unit regular vector field will undergo a rotation that is equal to the vertex angle defect. As the field is differentiable on triangles, the direction rotation accumulated along the cycle necessarily comes from direction discontinuities when crossing edges (Figure~\ref{fig:constflow}(b)). Such discontinuities can lead to the merging of streamlines on an edge where the flow outputs both adjacent triangles as in Figure~\ref{fig:constflow}(a). 

These issues were already addressed in \cite{gre2006}, where they define the field is defined on each vertex by a $2D$ vector in a local map, and interpolated on each triangle. In our case, we want to define a mapping from polyline entry point to polyline output point without numerical integration, making it useful to interpolate the field only along edge (not inside triangles). Moreover, we prefer interpolate the field in polar coordinates instead of Cartesian coordinates. It allows to represent more complex fields (with rotation greater than $\pi$ along an edge), and to better control field singularities by explicitly placing them on vertices. 

We represent the input vector field on each triangle by sampling the field direction at each edge extremity: $\alpha_k, k\in [0\dots 5]$ are the angles of the field, with respect to a reference vector $\overrightarrow{r}$ taken in the triangle plane. Note that due to angle defect and singularities on vertex, each vertex is associated to two angles: one for each incident edge. 

\begin{figure}[t]
\centerline{
\includegraphics[width=5.5cm]{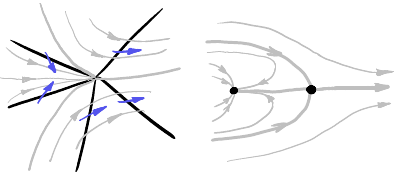}
\includegraphics[width=2.5cm]{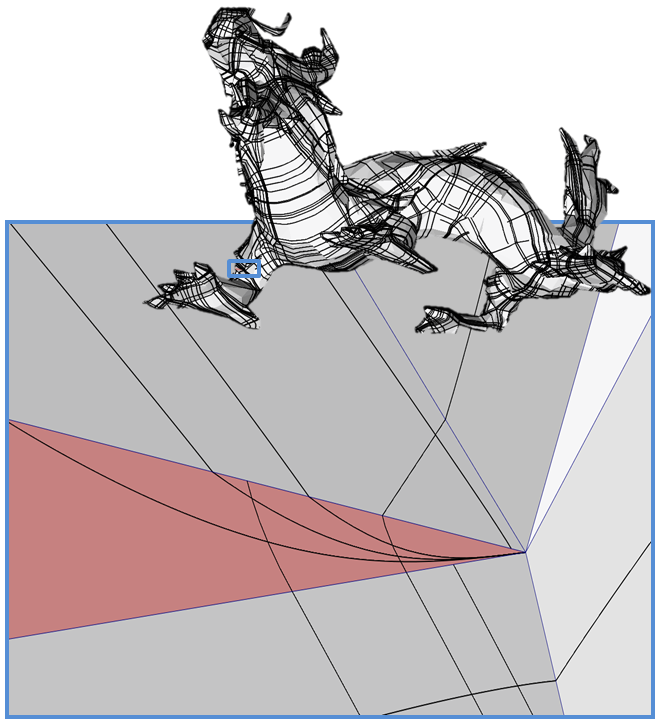}
}
\caption{When the field rotation is not homogeneous around a vertex, complex configurations can be represented: the vertex acts as a sink on the leftmost triangle, but as a regular vertex for other triangles (left). The continuous equivalent (middle) is a sink and a saddle, but located on a single vertex. An example is given on the right front leg of the dragon (right).}
 \label{fig:constrot}
\end{figure}

To prevent crossings, the input field must be continuous across edges i.e have the same angle with respect to an edge on both adjacent triangles of this edge. Another constraint we enforce is to evenly distribute, around each vertex, the angle discontinuity on triangle corners. This later constraint allows to prevent degeneracy (Figure~\ref{fig:constrot}) and to better manage singularities (Appendix~\ref{sec:streamlineExtremityOnVertex}).

\section{Stream-mesh}
\label{sec:algorithm}

A stream-mesh is the combinatorial representation of the field behavior inside a triangle of the mesh. It is a halfedge data structure decorated with additional information that represents the field. The field direction is given at each stream-vertex by its angle $\alpha$ relative to the triangle reference vector $\overrightarrow{r}$. Along each stream-halfedge $e$, the field have a unique behavior that may be :

\begin{itemize}
\item incoming ($I$) if the field points inwards the stream-face,
\item outgoing ($O$) if the field points outwards the stream-face,
\item tangent in the forward direction ($T_f$) if the field has the stream-halfedge direction,
\item or tangent in the backward direction ($T_b$) if the field direction is opposite to the stream-halfedge direction.
\end{itemize}

In this representation, we can define:
\begin{itemize}

\item An \textbf{in-list} as a list of stream-halfedges that contains at least one incoming stream-halfedge, and no outgoing stream-halfedge.
\item An \textbf{out-list} as a list of stream-halfedges that contains at least one outgoing stream-halfedge, and no incoming stream-halfedge.
\item A \textbf{simple stream-face} as a stream-face having a border that can be decomposed into an in-list, followed by a forward tangent stream-halfedge, followed by an out-list, and followed by a backward tangent stream-halfedge (Figure~\ref{fig:splitexple}--right).
\end{itemize}

The stream-mesh is initialized as a single stream-face by decomposing the triangle border according to the field behavior (Section~\ref{sec:tr2meshinit}). The main stream-face is then decomposed into simple stream-faces by a strategy inspired from the ear clipping algorithm \cite{earcut}: simple stream-faces are iteratively removed from the main stream-face until the main stream-face becomes simple (Section~\ref{sec:tr2meshsplit}).

\subsection{Main stream-face initialization}
\label{sec:tr2meshinit}

The initialization of the main stream-face from a triangle is performed independently on each interval between pairs of field samples. Each interval corresponds either to a triangle edge, or to a corner of the triangle between edge and next edge around the triangle.

\begin{figure}[t]
\centerline{\includegraphics[width=\linewidth]{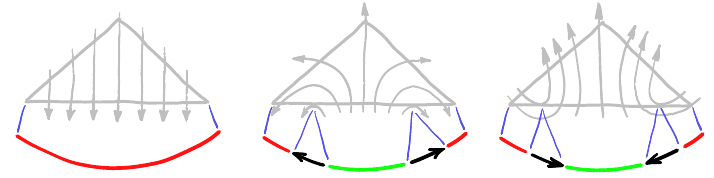}}
\centerline{\includegraphics[width=\linewidth]{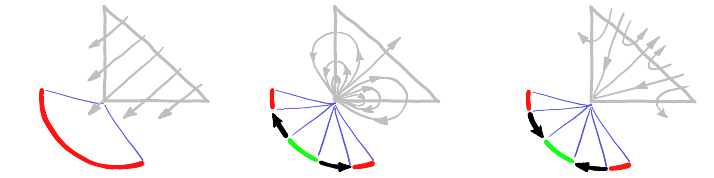}}
\caption{Combinatorial representation of the flow behavior. The first row shows the decomposition of an edge into incoming (green), outgoing (red), tangent forward and backward (black arrows), for three different fields. The second row shows that similar situations can occur on a triangle corner and can be characterized the same way. The field behavior is the same on both rows, but in the second row, the field rotation is performed on a single point instead of a triangle edge. The only difference between columns $2$ and $3$ is the tangent direction.}
  \label{fig:totopo}
\end{figure}

For the $k^{th}$ edge $E_k$ of the triangle, the angle of the field with respect to the edge is given by a linear interpolation between $\alpha_{2k}- \angle(\overrightarrow{r},\overrightarrow{E_k})$ and $\alpha_{2k+1} - \angle(\overrightarrow{r},\overrightarrow{E_k}) $. When this angle is equal to $0\mod 2\pi$ it is a forward tangent, when it is equal to $\pi\mod 2\pi$ it is a backward tangent, when it is strictly between $0$ and $\pi$ $\mod 2\pi$, it is incoming, and outgoing otherwise. A stream-halfedge is generated for every interval with constant type of behavior, including zero length intervals when the field is tangent at a single point.  As illustrated in the second and third columns of the top of Figure~\ref{fig:totopo}, tangents are required to characterize the field behavior.

On the triangle corner between $k^{th}$ edge $E_k$ and $j^{th}$ edge $E_j$ (with $j-k=1 \mod 3$), $\alpha_{2k+1}$ and $\alpha_{2j}$ may be different due to vertex angle defect or field singularities. As a consequence, it is possible for a vertex to contain important topologic information about the field. As illustrates Figure~\ref{fig:totopo}, the field behavior on a vertex (second row) is similar to its behavior along an edge (first row), and can be characterized in the same way. The segmentation is performed with the algorithm described for edges, but angles are linearly interpolated between $\alpha_{2k+1} - \angle(\overrightarrow{r},\overrightarrow{E_k})$ and $\alpha_{2j} - (\angle(\overrightarrow{r},\overrightarrow{E_k})+ \angle(\overrightarrow{E_k},\overrightarrow{E_j})) $. One can notice that using $\angle(\overrightarrow{r},\overrightarrow{E_k})+ \angle(\overrightarrow{E_k},\overrightarrow{E_j})$ instead of $\angle(\overrightarrow{r},\overrightarrow{E_j})$ allows to consider that the triangle border rotation on the corner is in $]0,\pi[$ (not modulo $2\pi$).
 
A possible geometric interpretation of stream-halfedges generated on triangle corners could be to consider the triangle as a rounded triangle having its corner radius tending to $0$. It makes the field and the triangle border rotating along the arc of circle instead of a single point.

\subsection{Split the stream-mesh into simple stream-faces}
\label{sec:tr2meshsplit}

\begin{figure}[t]
\reflectbox{\centerline{\includegraphics[width=\linewidth]{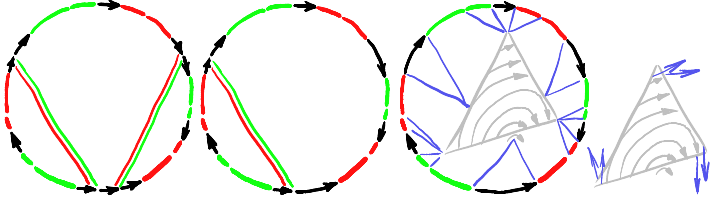}}}
\caption{The field is converted into a stream-mesh, then a simple stream-face is removed at each step until the main stream-face become simple.}
  \label{fig:splitexple}
\end{figure}

The stream-mesh is now initialized by a main stream-face. The decomposition iteratively removes a simple stream-face from the main stream-face until the main stream-face becomes simple (Figure~\ref{fig:splitexple}).

\begin{figure}[t]
%\centerline{\includegraphics[width=4cm]{splitrule.png}}
\centerline{\includegraphics[width=\linewidth]{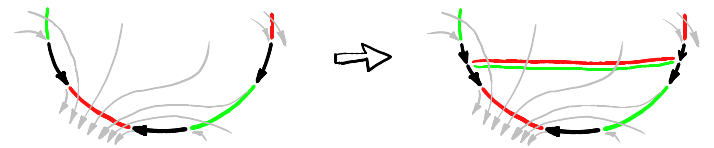}}
\caption{Splitting the main stream-face (left) by our rule produces a simple stream-face and removes a pair of in-list/out-list of the main stream-face border.}
  \label{fig:splitrule}
\end{figure}

To remove a simple stream-face (Figure~\ref{fig:splitrule}), we search in the stream-halfedges list of the main stream-face a sequence of edges that can be decomposed into : a $T_f$ stream-halfedge, followed by an out-list, followed by a $T_b$ stream-halfedge, followed by an in-list, and followed by a $T_b$ stream-halfedge. We split the first $T_f$ and last $T_b$ stream-halfedges of the sequence and introduce a new stream-edge linking the stream-vertices produced by the stream-edge split. The type of produced stream-halfedges is set to incoming in the simple stream-face side, and outgoing in the main stream-face side.

As illustrated by Figure~\ref{fig:splitrule}, the type of produced stream-halfedges is coherent with the flux that can be computed across the stream-halfedge. Indeed, the triangle border being convex, the field direction at the new stream-halfedge extremities will always point to the same half-plane of the new stream-halfedge. Moreover, the removed sequence prevents high rotations of the field on the stream-halfedge that would make it pointing in the other half-plane inside the stream-halfedge.

By symmetry, it is also possible to apply the same operation on the opposite field i.e. replace both $T_f \Leftrightarrow T_b$ and in-list$\Leftrightarrow$ out-list in the pattern and in the result.

Recursively applying the split operation converges to a decomposition into simple stream-faces, as demonstrated in Appendix~\ref{sec:convergence}.

\section{Streamline tracing algorithm}
\label{sec:tracingalgo}

\setlength\columnsep{.7\marginparsep}
\setlength\intextsep{.3\marginparsep}
\begin{wrapfigure}{r}{.3\linewidth}
\reflectbox{\includegraphics[width=\linewidth]{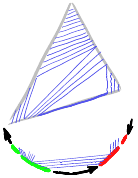}}
%\caption{Valid half-edge structure of a geometrically overlapping streamlines.}
% \label{fig:overlap}
\end{wrapfigure}

Streamlines are traced by iteratively traversing each triangle until it reaches a sink vertex or the surface boundary. Each triangle is traversed by iteratively traversing each simple stream-face (Figure \ref{fig:crossmesh}) by the algorithm detailed in this section. As illustrated by the inset figure, having polyline's segment inside each triangle results in: some triangular regions of the triangle that are not covered by polylines, and possible local geometric overlaps (on bottom edge). If polylines are dedicated to segment the mesh, such overlaps will result in faces with degenerated geometry, but it does not affect the combinatorics of the segmented mesh.

\subsection{Crossing a simple stream-face}
\label{sec:crosscell}

Crossing a simple stream-face requires defining how points of the in-list are mapped to points of the out-list. Any such mapping that does not cross streamlines will produce globally cross-free streamlines. However, it is better to choose a mapping that preserves as much as possible the field geometry. Our mapping is defined such that if an evenly distributed set of streamlines enters the triangle, it will leave it with an even distribution, except if field sinks or streamlines that are tangent to the boundary prevents it. It can be restated as follows: for a normalized field, if the stream-face is split by a streamline, both parts should have the same ratio between inflow and outflow. Here, we call by flux the amount of streamlines outgoing from a portion of the out-list (and symmetrically for the in-list). However, it can be considered as an abuse of terminology since an infinite set of streamlines may leave the triangle in a sink vertex, where the flux should be null. This heuristic perfectly respects the field when it is constant inside the triangle, and is evaluated in Section~\ref{sect:stresstest} in more difficult situations.

\begin{figure}[t]
\centerline{\includegraphics[width=\linewidth]{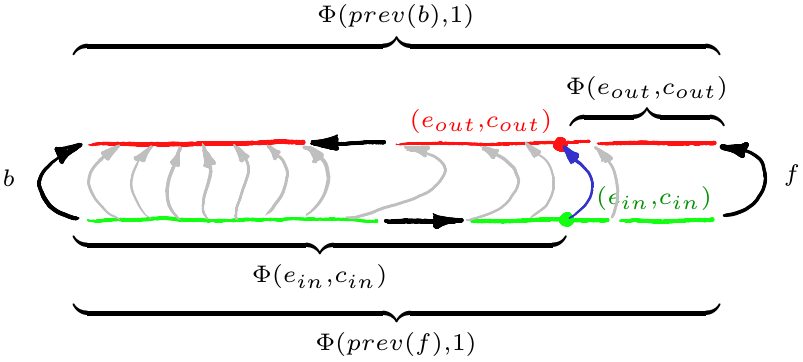}}
\caption{Important flows used to cross a simple stream-face.}
 \label{fig:flownotation}
\end{figure}

As illustrated in Figure~\ref{fig:flownotation}, we call $f$ (resp. $b$) the stream-halfedge of type $T_f$ (resp. $T_b$) that comes after the out-list (resp. before the in-list). 

We denote by $\Phi(e,c)$ the flux crossing the in-list (resp.out) of stream-halfedges up to the point located at the $(c,1-c)$ barycentric coordinate on the stream-halfedge $e$. It is recursively defined by $\Phi(e,c) = \Phi(prev(e),1) + \phi_{e}(c)$ where $\Phi(f,1) = 0$,$\Phi(b,1) = 0$, and $\phi_{e}(c)$ is the flux crossing the stream-halfedge $e$ up to the point of barycentric coordinates $c,1-c$. Computing $\phi_{e}(c)$ and its inverse are detailed in sections \ref{app:flow} and \ref{sec:solve}.

Using these notations (Figure~\ref{fig:flownotation}), the condition for a streamline to split the simple stream-face into two stream-faces having the same ratio between inflow and outflow writes:
\begin{equation}
\frac{\Phi(e_{in},c_{in})}{\Phi(prev(f),1)} = 1 - \frac{\Phi(e_{out},c_{out})}{\Phi(prev(b),1)}
\nonumber
\end{equation}
 
where the input point is $e_{in},c_{in}$ and the output point is  $e_{out},c_{out}$. As a consequence, the output point is given by :

\begin{equation}
(e_{out},c_{out}) = \Phi^{-1}\left(\Phi(prev(b),1) \left( 1-\frac{\Phi(e_{in},c_{in})}{\Phi(prev(f),1)}\right)\right)
\nonumber
\end{equation}

To compute the output position $({e}_{out},c_{out})$ of a streamline, we need to evaluate the functions $\Phi$, and $\Phi^{-1}$. The function $\Phi$ can be evaluated from $\phi_{e}(c)$ using its recursive definition. The function $\Phi^{-1}(x)$ requires to take the stream-halfedge $e$ such that $\Phi(e,0) \leq x \leq \Phi(e,1)$ and $\phi_{e}(1) \neq 0$, and to define its barycentric coordinate $c = \phi_{e}^{-1}(x-\Phi(e,0))$. 

As a consequence, we only need to be able to evaluate $\phi_{e}(c)$ and its inverse $\phi_{e}^{-1}(x)$ to cross a simple stream-face. 

\subsection{Computing $\phi_{e}(c)$}
\label{app:flow}

On edges, we set $\phi_{e}(c)$ to be the flux of the normalized vector field across the stream-halfedge $e$. The flux across the stream-halfedge is:

\begin{eqnarray}
\label{eq:flowonhalfedge}
\phi_{e}(c) &=& |\overrightarrow{e}| \int\limits_{0}^{c}{-\sin (\alpha_{o}+ t(\alpha_{d}-\alpha_{o})-\angle(\overrightarrow{e},\overrightarrow{r}))dt}\nonumber
\\
 &=& |\overrightarrow{e}|  \frac{\cos(\alpha_{o}+ t(\alpha_{d}-\alpha_{o})-\angle(\overrightarrow{e},\overrightarrow{r}))}{\alpha_{d}-\alpha_{o}}\Big|^c_0 \nonumber
\end{eqnarray}

where $\alpha_{o}$ and $\alpha_{d}$ are the field directions located at the vertex pointed by the stream halfedges $prev(e)$ and $e$, and expressed by their angle relative to $\overrightarrow{r}$ .

On corners, we can generally say that there is no flux that leaves the triangle i.e. $\phi_{e}(c) = 0$. However, for singularities with positive index such as source and sinks, there is an infinity of streamlines that reach or start from the corner (Figure~\ref{fig:singularity}). If an outflow stream-halfedge $e$ is defined in a triangle corner, in a sequence $T_f,O,T_b$, then we set $\phi_{e}(c) = c$. By symmetry,  if an inflow stream-halfedge $e$ is defined in a triangle corner, in a sequence $T_b,I,T_f$, then we set $\phi_{e}= -c$. This strategy provides a field behavior coherent with the continuous behavior of streamlines on field singularities as explained in appendix \ref{sec:streamlineExtremityOnVertex}.

\subsection{Computing $\phi_{e}^{-1}(x)$}
\label{sec:solve}

Computing $\phi_{e}^{-1}(x)$ requires to invert Equation~\eqref{eq:flowonhalfedge}. As cosine is not a one to one function, determining $\phi_{e}^{-1}(x)$ requires to take into account that it is a barycentric coordinate in the halfedge $e$, and therefore $0 \leq \phi_{e}^{-1}(x) \leq 1$. This constraint fixes $s \in \{-1,1\}$ and $k \in \mathbb{Z}$ in the formula:

\begin{eqnarray}
\phi_{e}^{-1}(x) &= &\frac{
s \arccos(\cos(\alpha_{o} - \angle(\overrightarrow{e},\overrightarrow{r}_T)) - x \frac{(\alpha_{d}-\alpha_{o})  }{ |\overrightarrow{e}|})
}{\alpha_{d}-\alpha_{o}} \nonumber \\
&+& \frac{
2k\pi- \alpha_{o} + \angle(\overrightarrow{e},\overrightarrow{r}_T)
}
{\alpha_{d}-\alpha_{o}}\nonumber 
\nonumber
\end{eqnarray}

\section{Discussion}
\label{sec:discussion}

This section evaluates the performances and robustness of our algorithm on synthetic stress tests (Section \ref{sect:stresstest}), proposes some applications where tracing robust streamlines is required (Section \ref{sec:application}), and provides some implementations details (Section \ref{sect:implementation}).

\subsection{Synthetic tests}
\label{sect:stresstest}

\begin{figure}[t]
\centerline{
\includegraphics[width=\linewidth]{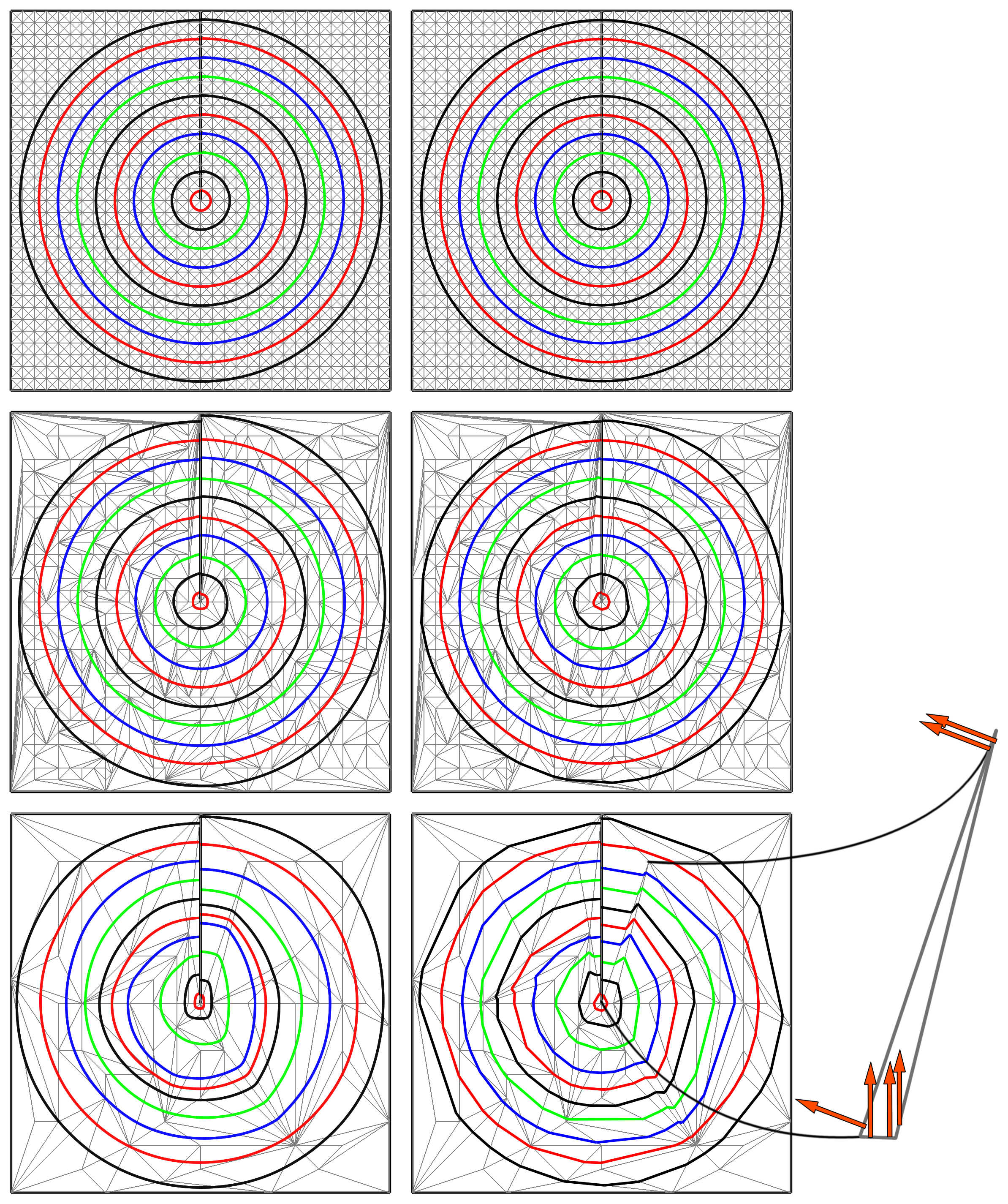}
}
\caption{Our algorithm on mesh (right column) is compared with a numerical integration on the same data (left column), with decreasing mesh quality from top to bottom.} 
  \label{fig:circle}
\end{figure}

To evaluate the geometric quality of our polylines, we traced them on a circular vector field with different mesh quality (Figure~\ref{fig:circle}). It shows our polylines smoothness and accuracy with different triangle qualities (upper to lower) and different field rotation magnitude (border to center). In practice, computer graphic meshes are closer to the upper and middle images, and field design algorithms tends to produce as smooth as possible fields. It is interesting to notice that an important loss of accuracy only appears (close-up) on very stretched triangles like one having a corner with a field singularity we zoomed on.

The robustness of our algorithm with respect to the mesh geometry is tested in Figure~\ref{fig:teaser}(left). No streamlines cross each other, and the loss of quality of the distribution is mostly due to the field smoothing algorithm used to generate the field.

\subsection{Applications}
\label{sec:application}

\begin{figure}[t]
\centerline{\includegraphics[width=.35\linewidth]{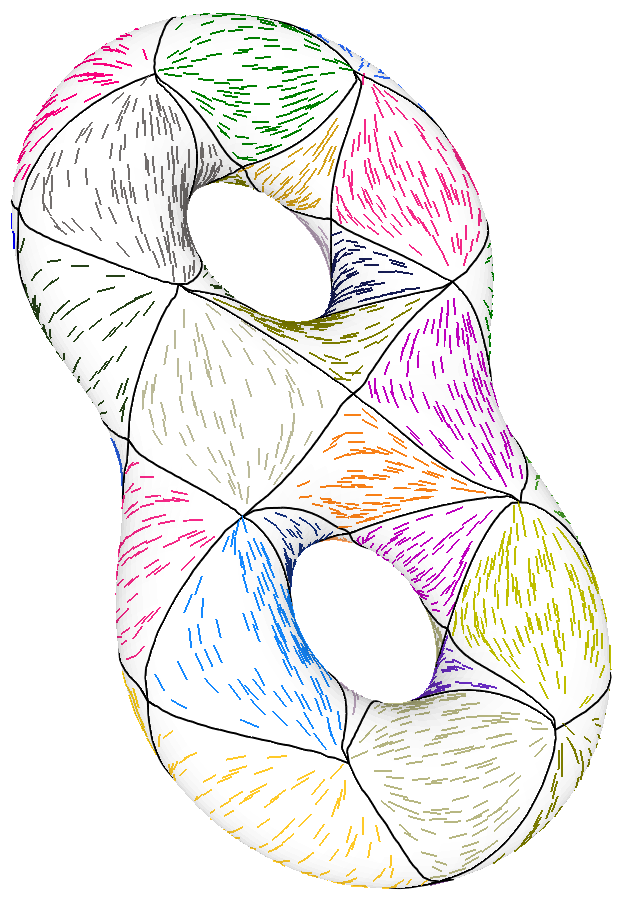}\includegraphics[width=.6\linewidth]{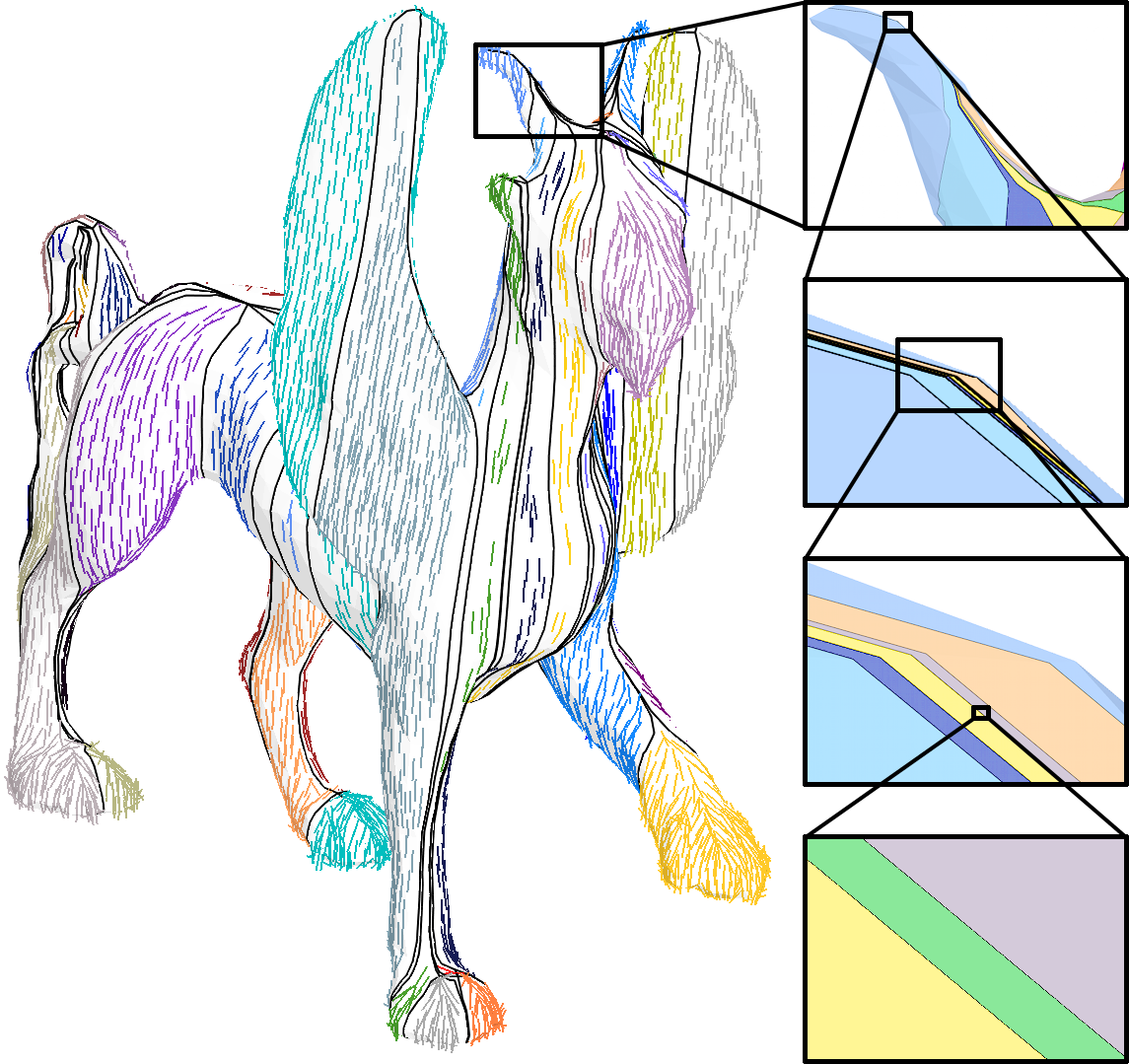}}

\caption{Morse-Smale complexes are computed using a Laplacian eigenfunction for the double torus and the $z$ coordinate for the Feline. Close-ups allow to see that polylines can be very close to each other without merging. The green chart visible in the upper close-up requires an important zoom factor (bottom close-up) to be noticed at the top of the horn.}
  \label{fig:gregre}
\end{figure}

\begin{figure}[t]
\centerline{\includegraphics[width=\linewidth]{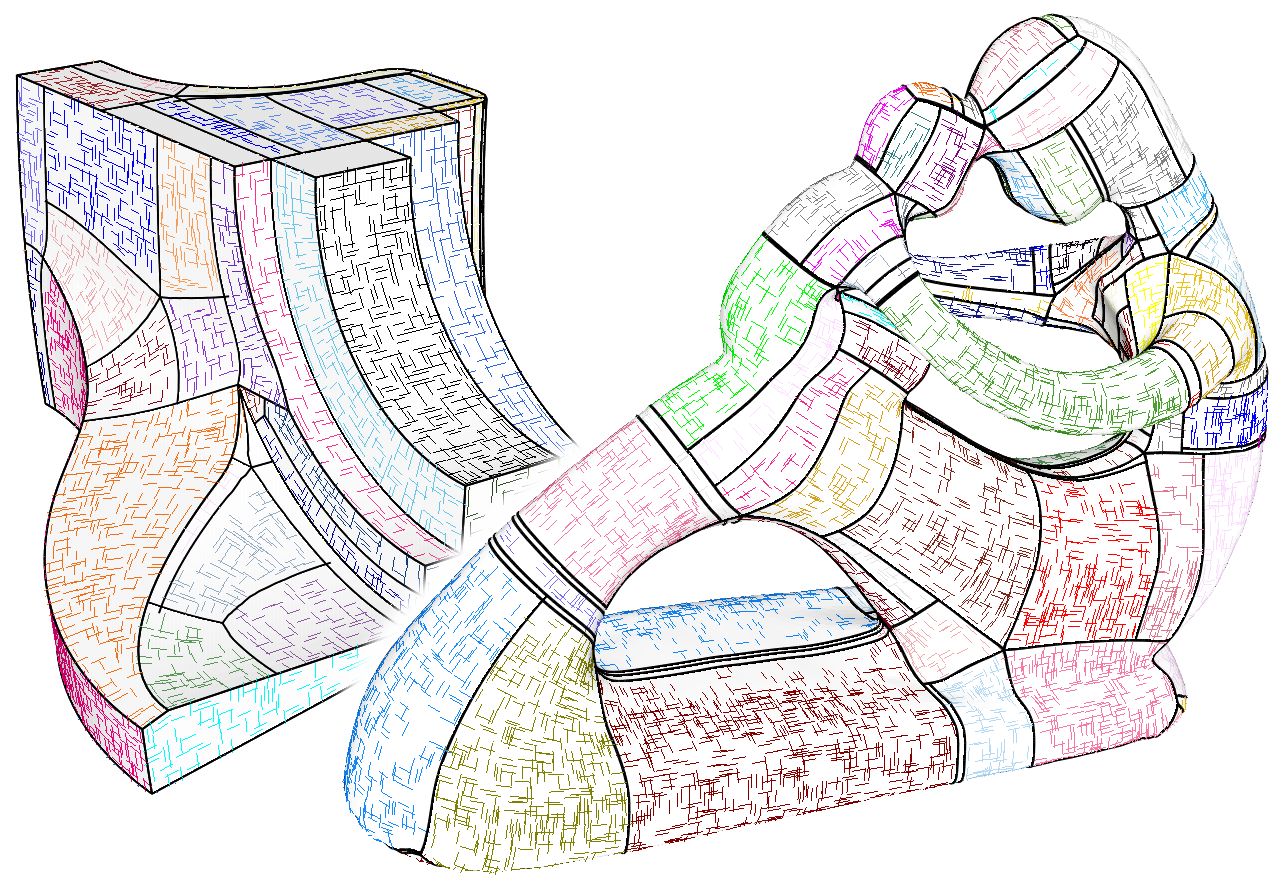}}
\caption{Tracing streamlines (black curves) from singularities of a cross field provides a decomposition of the surface}
  \label{fig:tmesh}
\end{figure}

We illustrate two possible applications of our method: computing Morse-Smale complexes (Figure~\ref{fig:gregre}), and splitting a mesh according to a direction field. Tracing streamlines of a N-symmetry direction field \cite{Kowalski2013} allows to partition $2D$ meshes. To illustrate a possible application of our method, we applied the same strategy on $3D$ surfaces, by growing all streamlines simultaneously, and stopping them when they reach a streamline defined on a perpendicular direction. As a result (Figure~\ref{fig:tmesh}) we obtain quadrangular charts with T-junctions everywhere except when a degeneracy is prescribed by feature curves as in the fandisk model. Such T-meshes could be useful after optimization, as proposed in \cite{tmesh}.

\subsection{Implementation notes}
\label{sect:implementation}

In our implementation, angles are represented by floating points. A direct consequence is that the border mapping function is one to one only up to numerical precision. More importantly, it could also produce inconsistencies along an edge if its decomposition into stream-halfedges does not match on both adjacent triangles. To prevent streamlines to be stuck inside an edge due to this issue, we always apply the decomposition on the same halfedge, and revert it to obtain the decomposition of the opposite halfedge.

Another issue comes from applications where the field have to be tangent to an edge. Using the proposed representation, it requires to have $\alpha_{2k} = \angle(\overrightarrow{E_k},\overrightarrow{r}) \mod \pi$ which is usually impossible due to floating point representation. To overcome this difficulty, we represent angles $\alpha_{2k}$ and $\alpha_{2k+1}$ relative to the $\overrightarrow{E_k}$ instead of $\overrightarrow{r}$, and express it in degree (in $[0\dots 360]$) instead of radian (in $[0\dots 2\pi]$). Tangents fields are then defined by angles $0 \mod 180$ which can be exactly represented by floating points.

\section*{Conclusion}

Tracing intersection-free polylines makes it easier to design new algorithms inspired from the continuous settings. Possible improvements of the method include using polycurves inside triangles, or finding a simpler way to cross each triangle. The question of the generalization to higher dimension arises naturally, but it is important to remember that the main issue (angle defect) requires that the metric is not induced by the object itself (for surfaces, it is induced by its embedding in $3D$ space, but volumes in $3D$ do not have this issue).

\bibliographystyle{alpha}
\bibliography{robust_streamline_no_convert}

\appendix

\section{\textbf{Behavior on vertices}}
\label{sec:vertexbehavior}

\subsection{Vertex indices}

N-symmetry direction fields may have singularities that can be characterized by their index. The index is well defined for smooth manifolds \cite{Mrozek95conleyindex}, and has been extended to triangulated surfaces~\cite{DFD}. In our case, we assume that singularities can only appear on vertices, leading to the following characterization of indices:
\begin{equation}
\label{eq:indice}
Index(A) = \sum \frac{\Delta\alpha_e}{2\pi} + \frac{2\pi -\sum \beta_e}{2\pi}
\end{equation}

where the sums are performed on all triangle corners referred by their halfedge $e$ incident to $A$, $Index(A)$ is the index of vertex $A$, $\Delta\alpha_e$ is the angle discontinuity on the triangle corner, $\beta_e$ is the triangle corner angle. The first sum is the total amount of field rotation around $A$, and the rest is the angle defect of $A$ divided by $2\pi$. 

\begin{figure}[t]
\centerline{\includegraphics[width=\linewidth]{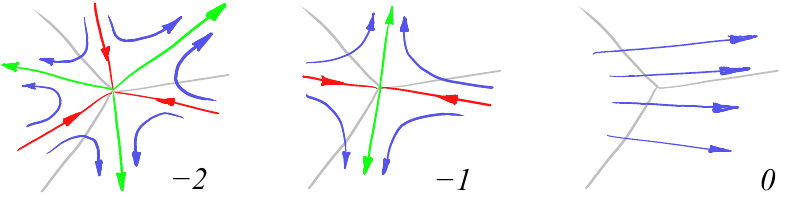}}
\centerline{\includegraphics[width=\linewidth]{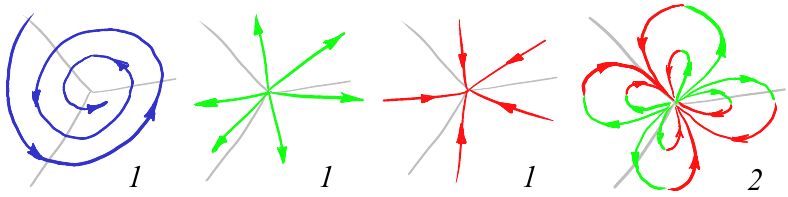}}
\caption{Singularities classified by index. On negative indices, there exist a finite number of streamlines (red and green) having the vertex as extremity. On regular vertices (index is zero), at most one streamline can cross the vertex. On positive index singularities, there exist an infinity of streamlines having the vertex as extremity, expect for the vortex case (lower-left).}
 \label{fig:singularity}
\end{figure}

Examples of singular vertices are given in Figure~\ref{fig:singularity}. One can notice that an infinite number of streamlines can reach the vertex only for strictly positive indices, leading to two different behaviors of our algorithm as detailed bellow.

\subsection{Geometric vertex crossing}

\begin{wrapfigure}[5]{r}{.2\linewidth}
\includegraphics[width=\linewidth]{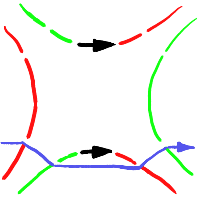}
\end{wrapfigure}

The default behavior of our algorithm is when there is not an infinity of streamlines having the vertex as one of its extremities. In this case, when a streamline leaves a triangle on a vertex location, the output of the triangle crossing algorithm is an adjacent edge, with a barycentric coordinate being either $0$ or $1$ to fit the vertex location. The streamline then continues on the next triangle until it ends in the vertex or leaves the vertex location as illustrated in the small figure to the right.

Our algorithm has this behavior because the flux on a stream-halfedge defined on a triangle corner is generally zero, and the constraint that the simple stream-face crossing algorithm is not allowed to generate outputs on a stream-halfedge without flux.

\subsection{Streamline extremity on a vertex}
\label{sec:streamlineExtremityOnVertex}

Streamlines may also have one of its extremities located on a vertex, but this occurs only for vertices with strictly positive index, as illustrated in Figure~\ref{fig:singularity} (we consider that if a unique streamline reach the vertex it will cross it with the previous behavior). We explain here why our way to determine the flux on stream-halfedges inside triangle corners (Section~\ref{app:flow}) gives non zero flux only for vertices with strictly positive index.

As the rotation speed of the field around the vertex $A$ is constant, the difference of angle $\Delta \alpha_e $ is equal to the sum of such rotations around the vertex $A$ times the ratio of $\beta_e$ over the sum of triangle corner angles around $A$. Putting it together with equation \eqref{eq:indice}, with summation over all halfedges $e^\prime$ pointing to $A$ gives:

\begin{equation}
\Delta \alpha_e = \frac{ \beta_e}{\sum \beta_{e^\prime}} \left(2\pi (Index(A) - 1) + \sum\beta_{e^\prime}\right)
\nonumber
\end{equation}

so the variation of angle with respect to halfedges pointing to $A$ is 
\begin{equation}
\Delta \alpha_e - \beta_e =  \frac{2\pi\beta_e  }{\sum \beta_{e^\prime}}(Index(A) - 1)
\nonumber
\end{equation}

As a consequence, if $\Delta \alpha_e - \beta_e$ is strictly positive, the vertex index is greater or equal to $1$. Else, the index is strictly less than $1$. Note that for direction fields with rational indices, we are still able to distinguish between singularities with and without flux.

In our algorithm, the condition to associate some flux to output stream-halfedges (defined on a triangle corner) is that the stream-halfedge must be contained in a sequence $T_f O T_b$. It means that the field angle with respect to the triangle border increases at least by $\pi$. Since the corner is convex, we have $\beta_e<\pi$. As a consequence, our algorithm gives some flux only for stream-halfedges in triangle corners corresponding to a vertex with strictly positive index. The same thing occurs for the sequence $T_b I T_f$.

\subsection{\textbf{Starting a streamline from a vertex}}

For vertices that are the origin of a finite number of streamlines (negative or null index), it is possible to generate all streamlines by simply starting a streamline for each inflow stream-halfedge on adjacent triangle corners. This is especially important for tracing streamlines from saddle points, as it is required for computing Morse-Smale complexes.

\section{Correctness of the decomposition}
\label{sec:proof}
\subsection{Convergence}
\label{sec:convergence}
Given a stream-face with $n$ in-lists and $n$ out-lists, let us choose one out-list as a reference.
Any two adjacent lists $i$ and $i+1$ have a tangent between them, let us define a sequence of labels $\{t_i\}_{i=0}^\infty$ as
the label of tangent stream-halfedge incident to both lists $i$ and $i+1$.
Then we define a sequence of integers $\{a_i\}_{i=0}^{+\infty}$ as follows: 
\begin{align*}
a_0&=0\\
a_{2i+1}&=\begin{cases} a_{2i}+1 &\text{~if~} t_{2i+1}=T_f,\\
a_{2i}-1 &\text{~otherwise}.
\end{cases}\\
a_{2i+2}&=\begin{cases} a_{2i+1}+1 &\text{~if~} t_{2i+2}=T_b,\\
a_{2i+1}-1 &\text{~otherwise}.
\end{cases}
\end{align*}
The defined sequence $\{a_i\}$ is arithmetic quasiperiodic: $a_{i+2n} = a_i - 2$ and is continuous in the sense that $|a_{i+1}-a_i|=1$.
% AUVP dans le sens que n change au fil des iterations
A stream-face is simple if and only if the corresponding sequence $\{a_i\}$ is decreasing.
The splitting rule described in section~\ref{sec:tr2meshsplit} searches for a pattern (per period $2n$) $(2i+1, 2i, 2i-1, 2i)$ in the sequence $\{a_i\}$
and replaces it with a new one $(2i+1, 2i)$. In other words, the splitting rule removes one (per period) local minimum of the sequence $\{a_i\}$.
The symmetric rule replaces $(2i+2, 2i+1, 2i, 2i+1)$ with $(2i+2, 2i+1)$, again removing a local minimum.
If a stream face is not simple, the corresponding sequence has at least one local minima, moreover, the sequence decreases by 2 with each period and therefore it is possible to apply one of the splitting rules.
Both rules keep the continuity of the sequence, and the period is reduced by 2 with each iteration, leading to a final decomposition of the initial stream-face into a set of simple stream-faces.

\subsection{Non-nullity of flux through simple faces}

We show that each simple stream-face is traversed by some flux. To do so we demonstrate that the out-list (as well as in-list) of a simple stream-face have non-zero associated flux.

First of all, let us note that all stream-halfedges created by splitting rules have non-zero flux. 
Indeed, their length is not zero: it is easy to see that due to the linear interpolation between angle samples,
the sequence $\{a_i\}$ is monotonic inside triangle corners; however the splitting rule searches for a local minimum of the sequence. Therefore, it is not possible to create a simple face entirely contained in a triangle corner.

Now let us show that all simple faces have non-zero flux through them. Let us suppose that the out-list of a simple stream-face has a zero flux. All outflow stream-halfedges on triangle edges as well as outflow stream-halfedges corresponding to splits have non-zero flux, since their length is greater then zero. The only option for an out-list to have a zero flux is to be contained in a triangle corner and to have $T_b, O, T_f$ structure, as defined in section~\ref{app:flow}. However it means that the corresponding sequence $\{a_i\}$ is increasing on this out-list, and that contradicts the monotonicity of the sequence $\{a_i\}$ for simple faces. Therefore, there is no out-list in a simple face that does not have a flux through it. The same argument shows by symmetry that there is no in-list without flux through it. 

\end{document}